\documentclass[a4paper]{cas-sc}

\usepackage[numbers]{natbib}

\usepackage[utf8]{inputenc} 
\usepackage[T1]{fontenc}    
\usepackage{hyperref}       
\usepackage{url}            
\usepackage{booktabs}       
\usepackage{amsfonts}       
\usepackage{nicefrac}       
\usepackage{microtype}      
\usepackage{xcolor}         
\usepackage{graphicx} 
\usepackage{subcaption}
\usepackage{amsmath}
\usepackage{amssymb}
\usepackage{algorithm}
\usepackage{algpseudocode}
\usepackage{wrapfig}

\DeclareMathOperator{\sign}{sign}

\newcommand{\bxi}{\boldsymbol{\xi}}
\newcommand{\bsigma}{\boldsymbol{\sigma}}

\newcommand{\bs}{\boldsymbol{s}}

\newcommand{\mI}{m_\text{I}}
\newcommand{\mF}{m_\text{F}}

\def\tsc#1{\csdef{#1}{\textsc{\lowercase{#1}}\xspace}}
\tsc{WGM}
\tsc{QE}

\begin{document}
\let\WriteBookmarks\relax
\def\floatpagepagefraction{1}
\def\textpagefraction{.001}

\shorttitle{Daydreaming Hopfield Networks}    

\shortauthors{Serricchio et al.}  

\title [mode = title]{Daydreaming Hopfield Networks 
and their surprising effectiveness on correlated data}  



%

\author[1,5]{Ludovica Serricchio}
\author[1]{Dario Bocchi}
\author[1]{Claudio Chilin}
\author[2]{Raffaele Marino}
\author[1,3]{Matteo Negri}
\author[1,4]{Chiara Cammarota}
\author[1,3,4]{Federico Ricci-Tersenghi}

\affiliation[1]{organization={Physics Department, University of Rome ‘La Sapienza’},
            addressline={Piazzale Aldo Moro 5}, 
            city={Rome},
            postcode={00185}, 
            country={Italy}}
\affiliation[2]{organization={Physics Department, Università degli Studi di Firenze},
            addressline={Via Sansone 1}, 
            city={Firenze},
            postcode={50019}, 
            country={Italy}}

\affiliation[3]{organization={CNR-Nanotec Rome unit},
            addressline={Piazzale Aldo Moro 5}, 
            city={Rome},
            postcode={00185}, 
            country={Italy}}
            
\affiliation[4]{organization={INFN sezione di Roma1},
            addressline={Piazzale Aldo Moro 5}, 
            city={Rome},
            postcode={00185}, 
            country={Italy}}
\affiliation[5]{organization={Center for Life Nano $\&$ Neuro-Science, Istituto Italiano di Tecnologia 291}, addressline={Viale Regina Elena}, city={Rome}, postcode={00161}, country={Italy}}


\begin{abstract}
To improve the storage capacity of the Hopfield model, we develop a version of the dreaming algorithm that \emph{perpetually} reinforces the patterns to be stored (as in the Hebb rule), and erases the spurious memories (as in dreaming algorithms). For this reason, we called it \emph{Daydreaming}. Daydreaming is not destructive and it converges asymptotically to stationary retrieval maps. When trained on random uncorrelated examples, the model shows optimal performance in terms of the size of the basins of attraction of stored examples and the quality of reconstruction. We also train the Daydreaming algorithm on correlated data obtained via the random-features model and argue that it spontaneously exploits the correlations thus increasing even further the storage capacity and the size of the basins of attraction. Moreover, the Daydreaming algorithm is also able to stabilize the features hidden in the data. Finally, we test Daydreaming on the MNIST dataset and show that it still works surprisingly well, producing attractors that are close to unseen examples and class prototypes.
\end{abstract}




\maketitle


\section{Introduction}

Hopfield Networks \cite{Hopfield1982} are among the most well-known models for storing and retrieving memories in a neural network.
The original Hopfield model, based on the Hebb rule \cite{hebb1949organization}, is both analytically tractable \cite{amit1985storing,amit1987statistical} and biologically plausible in its dynamics \cite{amit1989modeling}. Indeed, the retrieval dynamics in the original Hopfield model is defined by the following updating rule
\begin{equation}
    s_i^{(t+1)} = \sign \left(\sum_{j=1}^N J_{ij} s_j^{(t)}\right)\;, \qquad J_{ij} = \frac1N \sum_{\mu = 1}^P \xi_i^\mu \xi_j^\mu\;,
\end{equation}
where the $N$ neurons take discrete values ($s_i=\pm1$), and the $P$ discrete patterns $\{\xi^\mu\}_{\mu=1,\ldots,N}$ are the memories stored in the network. The retrieval of a memory is successful if the spin dynamics converge to one of the stored patterns.

Although the number of patterns $P$ that can be stored and efficiently retrieved in the Hopfield model is extensive, i.e.\ linear in the number of neurons $N$, the storage capacity (or maximum load) is rather limited. Indeed memories can be efficiently retrieved only if $\alpha=P/N < \alpha_c \simeq0.138$ \cite{amit1987statistical}, that is the number of stored memories is just a small fraction of the number of neurons.
Given this limitation of the Hebb rule, it is crucial to find other rules that can improve the storage capacity of the Hopfield model.

To overcome such limitations, researchers have introduced new strategies. One such strategy involves changing the nature of interactions by considering many-body terms (i.e.\ multi-spin interactions). This approach significantly enhances the capacity, making it more than linear \cite{gardner1987multiconnected, krotov2016dense, agliari2023dense}, and in some cases even exponential \cite{demircigil2017model, ramsauer2020hopfield, lucibello2023exponential}. While these strategies have demonstrated powerful applications, we will not discuss them in this work, as we are primarily interested in maintaining the Hebbian-like pairwise interactions, that lead to the simple and biologically plausible dynamics shown above.

In the original Hopfield model, the coupling matrix $J$ is given by the Hebb rule and does not need to be learned\footnote{Alternatively one can say that couplings are learned on the very first step as soon as the $P$ patterns have been seen once}. Analogously there exist other and more efficient rules that produce a different coupling matrix $J$ again without the need to learn it through a dynamical learning process (we review these rules in Section~\ref{subsubsec:analytic_rules}). In our approach, instead, we assume no a priori information and we learn the coupling matrix from the data (i.e. the patterns) through a learning dynamical process. In this learning context, the patterns are also called examples.

Our work elaborates on a learning strategy of the Hopfield model called \emph{dreaming} (or \emph{unlearning}) \cite{Hopfield1983}. The terminology is inspired by the hypothesis that the human brain, during the REM sleep phase, selectively erases useless memories while reinforcing useful ones \cite{crick1983function}. In this context, the Hebb rule can be interpreted as the ``day'' phase where the memories to be stored are encoded in the synapses' strengths (i.e.\ in the couplings), while the dreaming procedure can be interpreted as the ``night'' phase where spurious memories are erased.

However, most versions of the dreaming iterative procedure studied so far in the literature encounter significant challenges. Indeed, repeating the unlearning steps excessively often leads to a point where, depending on the load $\alpha$, even desired memories start deteriorating. This phenomenon results in catastrophic forgetting, where eventually no memory can be retrieved anymore unless the number of dreaming iterations is fine-tuned \cite{van1990increasing, van1992unlearning}. Additionally, the known dreaming procedures struggle to handle efficiently correlated examples \cite{van1997hebbian}.

Inspired by the concept of reinforcing the memories studied in \cite{fachechi2019dreaming}, in this work we design an iterative learning procedure that can be iterated indefinitely, thus avoiding fine-tuning. Moreover, it operates independently of any assumptions regarding the structure and correlation in the data (i.e.\ the pattern to be stored), marking a significant advancement in the field. Finally, the procedure only depends on a single parameter and keeps the conceptual simplicity of the original dreaming procedure. We call it \emph{Daydreaming}, as it drops any distinction between night and day cycles.

It is worth stressing that the Daydreaming procedure is fully local, that is the learning of the coupling $J_{ij}$ only depends on the values of neurons $s_i$ and $s_j$ (as well as the values of the patterns in $i$ and $j$). This is an important observation, given that any biologically plausible learning procedure must be local, while methods based on the inversion of the pattern correlation matrix are not local \cite{personnaz1985information, kanter1987associative}.

We test Daydreaming on both uncorrelated and correlated data, the latter being generated via the so-called \emph{random-features} (or \emph{hidden-manifold}) model \cite{goldt2020modeling}. This class of models has been proposed to generate more realistic datasets, closer to those typically used in machine learning applications \cite{gerace2022probing,baldassi2022learning}; the random-features scheme has been used to improve linear models (see for example \cite{rahimi2007random,louart2018random, mei2022generalization}). Moreover, it has been shown in \cite{negri2023storage} that the Hopfield model storing such structured data has a richer phase diagram, with a new \emph{learning phase} where the model learns to store features instead of examples (see Section \ref{sec:rf_data} for a brief review).
Therefore, random-features examples offer a good preliminary playground to test the retrieval performances of our algorithm (see Section \ref{sec:results_synt}). Surprisingly, we find that Daydreaming produces bigger basins of attraction for correlated data, signaling that the algorithm can \emph{spontaneously} exploit correlations in the data. Coherently with these findings, we obtain strikingly good results also on the MNIST dataset (see Section \ref{sec:results_mnist}).

\subsection{Related works}

Since we stressed that Daydreaming can be iterated indefinitely, here we review the literature on alternative learning rules, arguing that rules that avoided the pitfall of ``dreaming too much'' showed other undesirable properties (the most common ones being the addition of non-local terms and the vanishing of the basins of attraction).

\subsubsection{Learning rules with analytical expressions}
\label{subsubsec:analytic_rules}

Several strategies to increase the storage capacity avoid completely the use of the iterative learning procedure and rely on analytical expressions for the coupling matrix $J$, in the same spirit as the original Hebb rule. 
Many of the rules that we discuss in the following paragraph have also been found to be the fixed points of iterative procedures (although such an iterative procedure does not need to be run in practice).

The pseudo-inverse rule \cite{personnaz1985information, kanter1987associative} was introduced to deal with correlated data: it introduces a non-local term that makes it possible to store all the examples without error, regardless of their correlation. It is non-local in the sense that it requires the inversion of an $N \times N$ matrix, a task that cannot be done locally on a single neuron, and requires the intervention of a \textit{deus ex machina} that knows the status of the entire network. Another problem is that this rule requires looking at all the training examples at the same time, meaning that if we want to store an additional example we should repeat the procedure from scratch. This rule reaches the maximal theoretical capacity of symmetric networks, namely $\alpha_c=1$ \cite{gardner1988space}, but it presents a severe drawback: as $\alpha\to1$  the radius of the basins of attraction goes to zero, making retrieval possible only in the trivial case of a noiseless initial condition.

In \cite{dotsenko1991replica, dotsenko1991statistical}, the authors define a learning rule that is an interpolation between the Hebb and the pseudo-inverse rule. This rule can also be seen as the continuous limit of a certain dreaming procedure: the idea is to stop before the limit where basins of attraction vanish, in the same spirit as stopping the dreaming iterations. This rule also has a local-update version that is inspired by the learning with noise strategy suggested by Gardner \cite{gardner1989training}. Note that this rule needs to be modified in case of correlated examples \cite{der1992modified}.

In \cite{fachechi2019dreaming}, the authors deal with the problem of dreaming indefinitely long by adding a reinforcement term in the energy. They have a parameter that interpolates between the Hebb rule and a pseudo-inverse-like rule, in the same spirit as \cite{dotsenko1991statistical, dotsenko1991replica}, but it is not known if it is possible to converge to this rule with a local iterative update of the coupling matrix.

\subsubsection{Iterative learning rules}

The locality of the learning rule is a key property to make the rule biologically plausible and efficient to implement. Several modifications of the original dreaming procedure have been developed over the years to impose such a locality property. Unfortunately, they maintain the locality at the cost of other problems.
For example, in \cite{plakhov1992modified,plakhov1994converging,plakhov1995convergent} the authors define a dreaming procedure that uses local fields instead of spin configurations. This local rule converges to the pseudo inverse rule for a large number of iterations \cite{kanter1987associative}, which means that it has the same problem of vanishing basins of attraction. 
Moreover, in \cite{nokura1996paramagnetic, nokura1996unlearning} it was shown that even by modifying the rule by subtracting states sampled from the Gibbs measure at high temperature, the learning dynamics converge to a coupling matrix of the pseudo-inverse family.
Finally, in \cite{parisi1986memory, Marinari2019} the authors describe a local dreaming procedure that can be iterated indefinitely long, constantly learning new examples at the cost of forgetting the earlier ones, which means that this method does not improve the capacity of the Hebb rule.

An example of an iterative but non-local rule can be found in  \cite{benedetti2023eigenvector}, where the authors subtract the largest eigenvector of the coupling matrix at each step of dreaming. They find that this procedure has the same performance as the classical unlearning, but has the great advantage of being much more transparent and analytically accessible.

Recently, more advanced dreaming variations, inspired by the perceptron rule, have been studied in \cite{benedetti2022supervised, benedetti2023training}. They seem to achieve optimal results but still require the fine-tuning of some parameters (similarly  to the fine-tuning of the number of dreaming iterations). Furthermore, in \cite{agliari2023regularization}, the authors explicitly linked the number of dreaming iterations to regularization hyperparameters, interpreting the whole model through the lens of machine learning. Notably, they also discuss ways to fix these hyperparameters a priori.

Daydreaming is closely related to the maximum likelihood principle, as its update rule resembles a way to satisfy a moment-matching condition: see for example \cite{mackay2003information}, where the ``day'' and ``night'' terms are explicitly identified, and also algorithms of the contrastive divergence family \cite{hinton2005contrastive}, that are commonly used to train Restricted Boltzmann Machines (for some reviews, see for example \cite{decelle2021restricted} or \cite{lecun2015deep}). In this spirit, some learning rules related to ours (but less effective) have been discussed in \cite{kojima1995capacity,baldassi2018inverse} for a fully-connected symmetric model and generalized to sparse and asymmetric models in \cite{braunstein2011inference,saglietti2018statistical}. The key difference of Daydreaming is that we sample spurious minima at zero temperature.

Also in \cite{poppel1987dynamical} the authors discuss an update rule that samples at zero temperature, but instead of initializing the dynamics at random they use noisy versions of the training examples. This choice appears to produce smaller basins of attraction than our Daydreaming algorithm.

\section{Model, Algorithm, and Data}
\label{sec:definition}

\subsection{Hopfield networks}

A Hopfield network is a recurrent neural network made of $N$ neurons $\{s_i\}_{i=1}^N$ that can be in the states $\pm 1$ depending on the signal that they receive from all the other neurons. At each time step $k$, every neuron is updated according to the rule
\begin{equation}
    s_i^{(k+1)} = \sign \left( \sum_{j=1}^N J_{ij} s_j^{(k)}\right),
    \label{eq:update_rule}
\end{equation}
where $J_{ij}$ is the matrix of synaptic weights. We consider a symmetric matrix ($J_{ij}=J_{ji}$) with zeros on the diagonal ($J_{ii}=0$). We perform asynchronous updates, meaning that, at each time step, we update all the neurons  in an arbitrary order. The dynamics stop when spin values reach a fixed point.

This model works as an associative memory if, when initialized on a noisy version of one of the $P$ examples $\{\bxi^{\mu} \}_{\mu=1}^P$, the model converges to the clean version of such an example. We are interested in finding, in an efficient way, the synaptic weights that maximize the number of retrievable examples and the mean initial distance that still allows for correct memory retrieval.

\subsection{Daydreaming algorithm}

Our algorithm aims to enhance the storage capacity of a Hopfield network by simultaneously ``dreaming away'' spurious memories and reinforcing the desirable ones.
The removal part operates as in the original dreaming procedure \cite{Hopfield1983}: at each step $u$, we initialize the network to a random configuration,  and we run the update rule in Eq.~(\ref{eq:update_rule}) until we reach a fixed point $\bsigma^{(u)}$. Then we modify the coupling matrix to increase the energy of the fixed point configuration $\bsigma^{(u)}$ just found. Simultaneously, we reinforce one of the memories, i.e. we choose an index $\mu(u)$ uniformly at random and we decrease the energy of $\bxi^{\mu(u)}$.
In formulas, the Daydreaming update rule reads
\begin{equation}
    J^{(u+1)}_{ij}=J^{(u)}_{ij}+\frac{1}{\tau N} \left(\xi_i^{\mu(u)} \xi_j^{\mu(u)}-\sigma_i^{(u)}\sigma_j^{(u)}\right)\; ,
    \label{eq:learning_update}
\end{equation}
where $\tau$ is a timescale parameter that acts as an inverse learning rate and the factor $1/N$ ensures that the rule works well in the limit of very large networks. Although the the update rule in Eq.~(\ref{eq:update_rule}) is invariant under a global rescaling of the coupling matrix $J$, we prefer to keep it well bounded, and so we normalize it every $N$ steps.

The learning rule in Eq.~(\ref{eq:learning_update}) needs to be equipped with an initial condition. We have tried several uninformed choices --- e.g. $J_{ij}=0$ or $J_{ij}\sim \mathcal{N}(0,1)$ being $\mathcal{N}(0,1)$ the normal distribution --- and all converge to the same asymptotic coupling matrix. Just to make the learning process faster, we initialize $J_{ij}$ according to the Hebb rule.
The complete pseudo-code is reported in Alg.~\ref{alg:Daydreaming}.

\begin{algorithm}[t]
\caption{Daydreaming learning algorithm}
\begin{algorithmic}
\Require examples $\{\bxi^{\mu} \}_{\mu=1}^P$
    \State $J_{ij} \gets \frac{1}{N} \sum_{\mu} \xi_i^{\mu} \xi_j^{\mu}$ \Comment{Initialization with the Hebb rule}
    \State $J_{ii} \gets 0$
    \For {$t=1,\dots,E$} \Comment{Do $E$ epochs}
        \For {$u=1,\dots,N$} \Comment{Do $N$ steps in each epoch}
            \State $\mu \gets \mathrm{Unif}(\{1,\dots,P\})$ \Comment{Pick an example at random}
            \State $\sigma_i \gets \mathrm{Unif}(\{+1,-1\}) $ \Comment{Initialize spins at random}
            \While {not converged}
                \State $\sigma_i \gets \sign (\sum_j J_{ij} \sigma_j)$ \Comment{Run the spin update dynamics}
            \EndWhile
            \State $J_{ij} \gets J_{ij} + \frac{1}{\tau N} (\xi_i^{\mu} \xi_j^{\mu}-\sigma_i\sigma_j)$ \Comment{Update the coupling matrix}
            \State $J_{ii} \gets 0$
        \EndFor
        \State $J_{ij} \gets J_{ij} / ||J||_2 $ \Comment{Normalize after each epoch}
    \EndFor
\end{algorithmic}
\label{alg:Daydreaming}
\end{algorithm}

\subsection{Datasets}
\label{sec:rf_data}

We will test the Daydreaming algorithm on different datasets to measure the capacity and the mean size of basins of attraction in the Hopfield model with coupling matrix $J$ trained with the learning rule in Eq.~(\ref{eq:learning_update}).

Apart from the random patterns dataset, where $\xi_i^\mu=\pm1$ with equal probability, and the MNIST dataset \cite{lecun1998gradient}, we are going to use also the random features model \cite{goldt2020modeling}, which is defined as follows: $P$ examples $\bxi^{\mu}$ are generated as superpositions of $D$ random features $\mathbf{f}^k \in \{-1, +1\}^N$, namely
\begin{equation}
        \xi_i^\mu = \sign \bigg( \sum_{k=1}^D c_k^\mu f_i^k \bigg)
\end{equation}
where $c_k^\mu \sim \mathcal{N}(0,1)$ and $f_{ki}\sim \mathrm{Unif}(\{+1,-1\})$.

This model, in addition to the usual load parameter $\alpha=P/N$, is also controlled by the parameter $\alpha_D=D/N$, which describes how strongly correlated the examples are: if $\alpha_D \gg 1$ the distribution of the examples converges to $\mathrm{Unif}(\{+1,-1\})$ and we get back a dataset of uncorrelated examples; while if $\alpha_D \lesssim 1$ the examples are correlated, and the task of storing them becomes more complicated, even if in principle the correlation among example could be exploited to increase the capacity. For instance, the storage capacity of the Hebb rule decreases with $\alpha_D$. In addition to the usual \emph{storage phase}, the Hopfield model with the Hebb rule for correlated examples also exhibits a \emph{learning phase}: if $\alpha > \alpha^*(\alpha_D)$, examples are unstable and features become locally stable fixed points of the dynamics  \cite{negri2023storage}.
Understanding whether this rich behavior of the Hopfield model is present also with coupling matrices different from the Hebbian one is an open and very interesting question.

\section{Results on synthetic data}
\label{sec:results_synt}

Given a generic Hopfield network with coupling matrix $J$ and assuming that the retrieval of the stored patterns is performed following the parallel update rule in Eq.~(\ref{eq:update_rule}), we need to quantify the capacity of the network and the mean size of the basins of attraction. We do this by computing the so-called \emph{retrieval map} as follows.

We  define the \emph{magnetization} $m^\mu$ (or overlap) of a configuration $\bs$ with a given example $\bxi^\mu$ as $m^\mu=\frac{1}{N} \sum_{i=1}^N \xi_i^\mu s_i$.
We initialize the network on a configuration that has initial magnetization $m_\mathrm{I}$ with an example, we run the dynamics in Eq.~(\ref{eq:update_rule}) until convergence, and then we measure the final magnetization $m_\mathrm{F}$ with the chosen example.
After repeating the experiment several times and averaging over all the examples, the resulting curves are called \emph{retrieval maps} and are shown in Figs. \ref{fig:results_uncorr} and \ref{fig:results}.

From the retrieval maps, one can read out a lot of useful information.
For example, when the network reaches its capacity, the stored patterns (at least some of them) become locally unstable, i.e.\ they are no longer fixed points of the spin update dynamics.
Looking at the retrieval map, one realizes that the learned examples are \emph{locally stable} if $m_\mathrm{I}=1$ corresponds to a $m_\mathrm{F}\simeq1$.
Moreover, the mean size of the basins of attraction can be estimated from the length of the plateau with $m_\mathrm{F}\simeq1$.
Indeed one is interested in understanding how distant from a typical pattern the starting configuration can be placed such that the spin update dynamics converge to the chosen pattern. 
Given that a perfect retrieval ($m_\text{F}=1$) is not required and also $m_\text{F}\simeq1$ is fine for the definition of the basin of attraction, we do not provide a precise and quantitative definition.
Nevertheless, looking at the retrieval maps in Figs. \ref{fig:results_uncorr} and \ref{fig:results}, the presence of the plateau with $m_\mathrm{F}\simeq1$ is very clear, and also its length can be estimated very reasonably (especially if we just need to compare two different networks to classify them according to the plateau length).

\subsection{Convergence}
\begin{figure}[t]
\centering
\includegraphics[width=0.49\textwidth,clip]{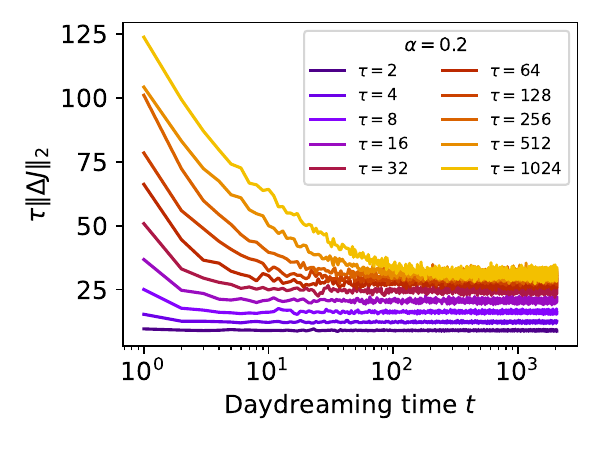}
\includegraphics[width=0.49\textwidth]{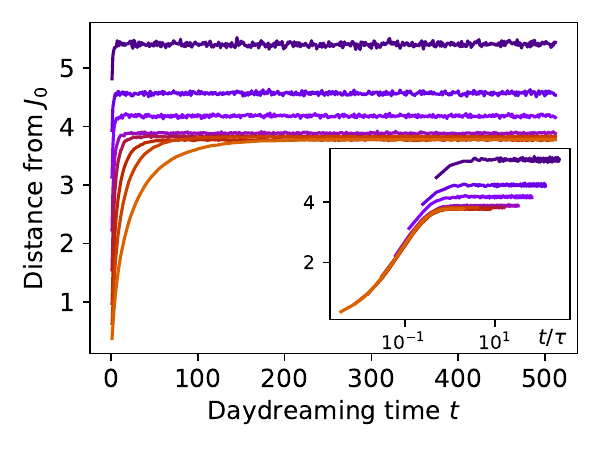}
\caption{ 
\textbf{The Daydreaming algorithm has a good asymptotic behaviour.} 
To illustrate the learning process of the coupling/synaptic matrix $J$, we plot, as a function of training time $t$, the rescaled norm of the increment  $\tau ||\Delta J||_2$ (left panel) and the distance of the coupling matrix $J$ from the Hebbian initial condition $J_0$ (right panel). The training process is very smooth, becomes stationary after a time $t\simeq \tau$ and,  for $\tau$ large enough, the asymptotic regime does not depend on the value of $\tau$. Inset: the curves as a fuction of $t/\tau$ collapse for $\tau \geq 64$. }
\label{fig:asymptotic}
\end{figure}

We start the analysis of the Daydreaming algorithm by studying its convergence toward the stationary asymptotic state. In Fig.~\ref{fig:asymptotic} we show the training of the coupling/synaptic matrix $J$ on uncorrelated data for $\alpha=0.2$ (a value larger than the capacity of the original Hopfield model with Hebbian matrix). It is worth stressing that the Daydreaming algorithm has just one parameter, $\tau$, which regulates the speed of training, and we are going to show that its precise value is not relevant as long as it is large enough.
So, in practice, Daydreaming does not need to be fine-tuned, and it is almost a parameter-free algorithm.

The training of the coupling  matrix $J$ is a dynamical stochastic process taking place in a very high-dimensional space\footnote{The number of independent elements in a $N\times N$ symmetric matrix with null diagonal elements is equal to $N(N-1)/2$.} and so we need to project it on simpler observables to study it.
In Fig.~\ref{fig:asymptotic} we plot two of such interesting observables: in the left panel we show the scaled norm of the matrix increment $\tau ||\Delta J||_2$, where  $\Delta J_{ij}^{(u)}=J_{ij}^{(u+1)}-J_{ij}^{(u)}$, and in the right panel the distance of the coupling matrix $J$ from the Hebbian initial condition $J_0$. The two observables provide complementary information: the latter measures the typical distance covered by the learning dynamics, i.e.\ how far it is from the starting point, while the former measures the typical size of the displacement per step.

Both quantities converge to a constant asymptotic value on timescales comparable with the rate $\tau$. We are not claiming that the matrix $J$ converges to a fixed point. Indeed, most of the time, the matrix $J$ keeps changing slowly, due to the stochastic nature of the learning rule, but maintains the same statistical properties, that eventually produce the same retrieval maps. In any case, the experiments shown in Fig.~\ref{fig:asymptotic} strongly suggest it is not useful to run the Daydreaming algorithm for a time much larger than $\tau$.

More importantly, the asymptotic state reached by Daydreaming slightly depends on $\tau$ only for very small values of $\tau$. As soon as $\tau$ is large enough (and a number of the order of a few tens of steps is enough) the results are $\tau$-independent, making the Daydreaming algorithm free from parameters to be tuned. The practical rule of choosing a reasonably large $\tau$ (e.g.\ $\tau \sim 100$) and running the algorithm for a time slightly larger than $\tau$ steps seems to work well and saves the time usually dedicated to the fine-tuning of algorithm parameters.

The great advantage of the Daydreaming algorithm is that it can be iterated indefinitely because the coupling matrix $J$ does not degrade over time and the corresponding Hopfield network maintains its retrieval capabilities, as shown from the convergence of the retrieval maps in panel (a) of Fig.~\ref{fig:results_uncorr} and in all panels of Fig.~\ref{fig:results}. 

For the above reason, the best signature for the convergence of the Daydreaming algorithm is the retrieval map being stationary in time. We are going to use this stronger criterion for all the different datasets studied in the present work (uncorrelated data, random-features data, and MNIST data).

\subsection{Retrieval of uncorrelated data} 

\begin{figure}[t]
\centering
\includegraphics[width=0.49\textwidth]{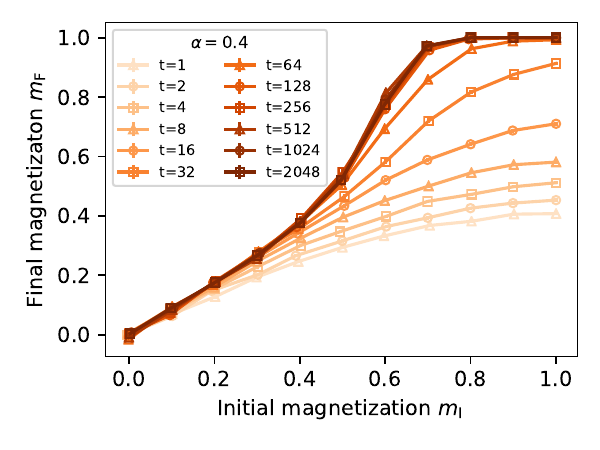}
\includegraphics[width=0.49\textwidth]{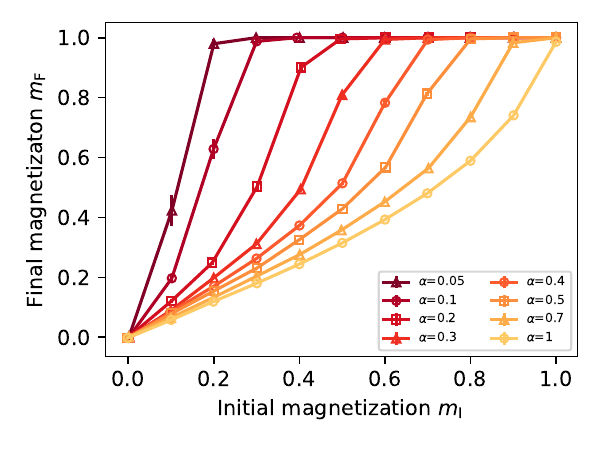}
\caption{ 
\textbf{The Daydreaming algorithm matches state-of-the-art performances in storing random data.} 
We report the retrieval maps for uncorrelated examples. In the left panel, we show the evolution of the retrieval map during the training with $\tau=256$ (from lighter to darker shades). The training converges around $t \simeq \tau$ and the asymptotic network has a basin of attractions consistent with the state-of-the-art results in \cite{benedetti2022supervised}. In the right panel, we show the asymptotic retrieval maps for several values of the load $\alpha$. A plateau with $m_\text{F}\simeq1$ exists for any $\alpha<\alpha_c=1$, i.e. the network trained with Daydreaming has a maximum capacity. We used $N=10^3$ for these figures.}
\label{fig:results_uncorr}
\end{figure}

In Fig.~\ref{fig:results_uncorr} we report the results obtained by running Daydreaming on uncorrelated data. In the left panel, we show the retrieval maps for $\alpha=0.4$ ($N=10^3$ is used in both panels). Different colors correspond to different training times, from lighter to darker.
At $t=0$ we start from the Hebbian matrix and indeed, for short times, the network fails in retrieving the random patterns stored, as can be deduced from the fact the curve in $m_\text{I}=1$ reaches values $m_\text{F}<1$.
This is expected given that $\alpha=0.4$ is larger than the capacity of the Hebbian coupling matrix ($\alpha_c\simeq0.138$).

During the training, the retrieval map improves, especially for large values of $\mI$, i.e. close to the stored patterns. Already for $t=128$ the stored patterns become locally stable and for $t\ge 256$ the retrieval map has reached its asymptotic shape, which would remain unchanged even keeping running the Daydreaming algorithm. The asymptotic shape is very interesting: first of all, it is much better than the one for Hebbian couplings, with a plateau extending down to $\mI\simeq 0.7$; moreover, the entire shape of the retrieval map matches the one found in Ref.~\cite{benedetti2022supervised}, which is considered optimal.
Surprisingly enough, without any fine-tuning, the Daydreaming algorithm spontaneously converges towards the optimal retrieval map. In other words, the Daydreaming algorithm can spontaneously modify the coupling matrix to accommodate a large number of patterns, maximizing the basin of attraction of these attractive fixed points for the Hopfield spin update dynamics.

The right panel of Fig.~\ref{fig:results_uncorr} shows the converged retrieval maps obtained at several values of $\alpha$, from very small values up to $\alpha=1$.
The stored patterns are stable up to $\alpha=1$ proving that the Daydreaming algorithm can reach the maximum capacity, $\alpha_c=1$ for uncorrelated patterns  \cite{gardner1988space}.
As expected, the size of the basins of attraction shrinks when $\alpha$ grows and becomes vanishing at $\alpha_c=1$.
Such a maximum capacity (with vanishing basins of attraction) has been already obtained with other algorithms implementing the dreaming procedure \cite{fachechi2019dreaming}.
However, the Daydreaming algorithm achieves these optimal performances by running a local (i.e. biologically plausible) learning rule, without the need for a rule involving the entire network (as when matrix inversions are required).

To understand how fast the Daydreaming algorithm learns the optimal retrieval maps shown in the right panel of Fig.~\ref{fig:results_uncorr}, the reader can look at the data plotted with red and reddish colors in panels (a), (b), and (c) of Fig.~\ref{fig:results}.
The convergence to the asymptotic state is faster when $\alpha$ is smaller and becomes slower approaching the maximum load.
It is worth noticing --- see panel (a) of Fig.~\ref{fig:results} --- that, even for low loads, when the Hebbian rule can store the pattern, the Daydreaming algorithm slightly improves over the Hebbian rule, by enlarging the basins of attraction.

Overall, we find that the Daydreaming algorithm matches the state-of-the-art results in retrieving uncorrelated examples in the whole $\alpha\in[0,1]$ range.

\subsection{Retrieval of correlated data}

\begin{figure}
\centering
\subcaptionbox{\hspace*{6cm}}{\includegraphics[trim={0.4cm 0.5cm 0.0cm 0.0cm},width=0.49\textwidth,clip]{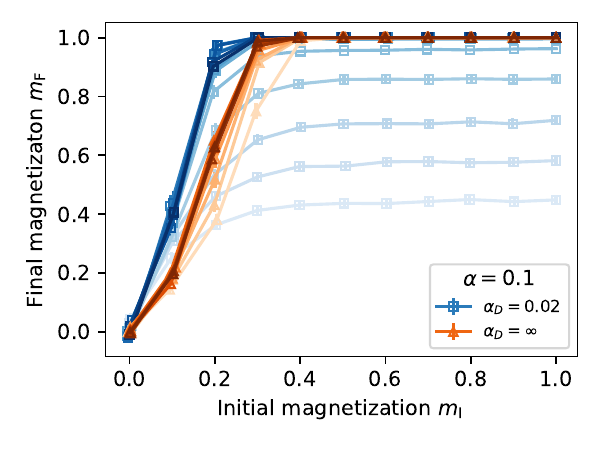}}
\subcaptionbox{\hspace*{6cm}}{\includegraphics[trim={0.4cm 0.5cm 0.0cm 0.0cm},width=0.49\textwidth,clip]{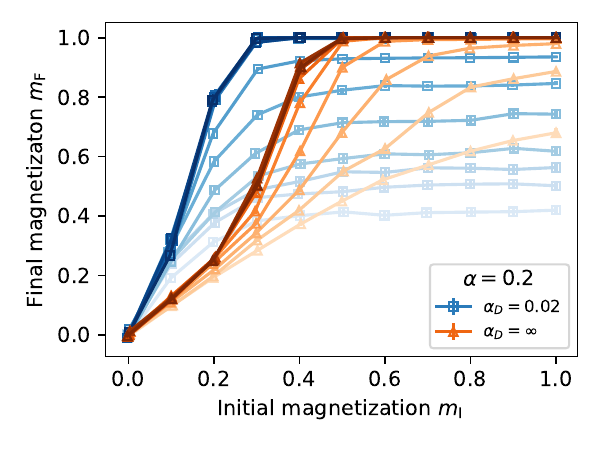}}
\subcaptionbox{\hspace*{6cm}}{\includegraphics[trim={0.4cm 0.5cm 0.0cm 0.0cm},width=0.49\textwidth,clip]{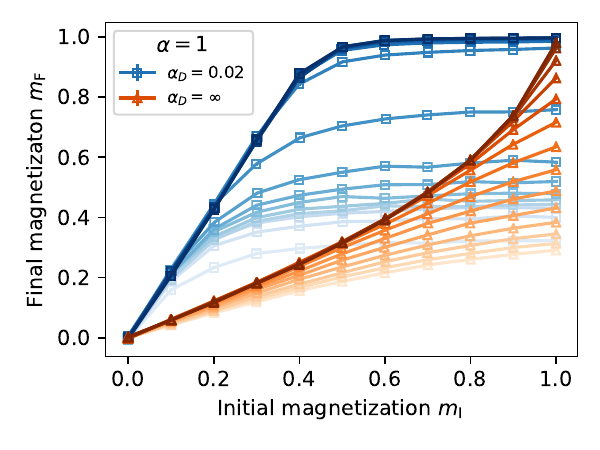}}
\subcaptionbox{\hspace*{6cm}}{\includegraphics[trim={0.4cm 0.5cm 0.0cm 0.0cm},width=0.49\textwidth]{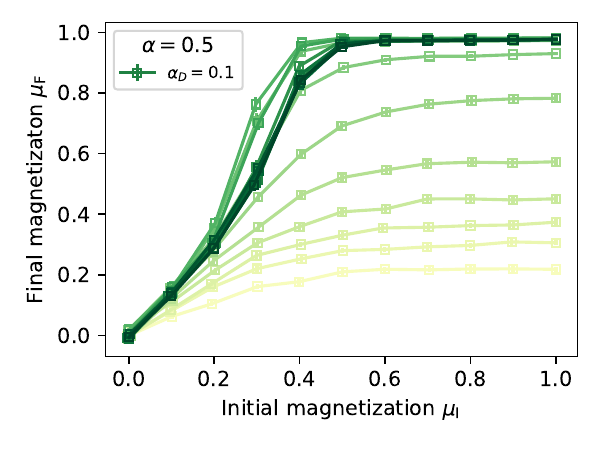}}
\caption{\textbf{For the Daydreaming algorithm, correlated examples are easier to store and retrieve than random ones}. We show the retrieval maps during the training process with the Daydreaming algorithm. In panels (a-c), blue squares are for correlated data, while red triangles are for uncorrelated data. Different shades of the colors represent different training times: the lightest color is $t=1$ and the darkest is $t=32768$ (logarithmic spacing). 
Panel (a) shows results for $\alpha<0.138$, where Daydreaming enlarges the basin of attraction of uncorrelated examples and stabilizes the correlated examples.
Panel (b) shows results for $\alpha>0.138$, where uncorrelated data become stable faster but correlated data end up with a larger basin of attraction.
Panel (c) shows results for $\alpha=1$, where uncorrelated data become stable attractors at the end of the training but with vanishing small basins of attraction; instead, correlated data have large basins of attraction.
Panel (d) shows the retrieval map for the features hidden in the data, that get stabilized as well by the Daydreaming algorithm.
We used $N=10^3$ for these figures.
}
\label{fig:results}
\end{figure}

In panels (a), (b), and (c) of Fig.~\ref{fig:results} we show how the retrieval map changes during the training with the Daydreaming algorithm for uncorrelated data (red triangles) and for correlated data generated by the hidden features model (blue squares). The plotted results have been obtained with $N=10^3$, and values of $\alpha$ and $\alpha_D$ shown in the legend. Different shades of the colors represent different training times: the lightest color is $t=1$ and the darkest is $t=32768$ (logarithmic spacing). We observe that the convergence is faster for smaller $\alpha$ values and slower for larger $\alpha$ values. In any case, the Daydreaming algorithm converges well and without the need for any fine-tuning of parameters, even when it is run at the network's maximum load ($\alpha=1$ in the case of uncorrelated examples).

Even for a low load --- see panel (a) of Fig.~\ref{fig:results}, where $\alpha=0.1$ --- we see a clear improvement of the Daydreaming algorithm over the Hebbian rule. For uncorrelated data, we observe the formation of a slightly larger basin of attraction. For correlated data, with the Hebbian coupling matrix, the examples are unstable and get spontaneously stabilized by the Daydreaming algorithm. At convergence, the correlated examples have large basins of attraction than uncorrelated pattern.

Panel (b) of Fig.~\ref{fig:results} shows results for a medium load ($\alpha=0.2$). In this case, the dreaming procedure is strictly required to have stable examples. Although the learning from uncorrelated examples converges faster, the coupling matrix $J$ trained on correlated examples eventually has larger basins of attraction. This is quite surprising and it means the Daydreaming algorithm has the right capabilities to extract useful correlations from a dataset. And it can do it by itself, without being specifically instructed.

Finally panel (c) of Fig.~\ref{fig:results} report results for the case of a large load ($\alpha=1$). Although the learning dynamic is slower, the Daydreaming algorithm converges to very good (probably optimal) retrieval maps. For uncorrelated data, the retrieval map reaches the point $(\mI=1,\mF=1)$ with a vanishing plateau, meaning we are running at the maximum load $\alpha_c=1$. Instead, the basin of attraction for correlated examples is still very large at $\alpha=1$, meaning that the retrieval will also be possible for $\alpha > 1$. This implies that the Hopfield model endowed with the Daydreaming training process can exploit the correlation to overcome the storage bound of uncorrelated examples.
Daydreaming extends the storage limit of correlated examples described in \cite{negri2023storage}, given that all the values of $\alpha$ and $\alpha_D$ reported in panels (a-c) of Fig.~\ref{fig:results} are outside the storage phase.

\subsection{Retrieval of features}
In \cite{negri2023storage} it has been shown that the Hebb rule applied to correlated data can stabilize (and thus retrieve) the hidden features in the so-called learning phase ($\alpha_d<0.138$ and $\alpha$ large enough).
A natural question is whether the same happens if the coupling matrix $J$ is learned via the Daydreaming algorithm.
To this end, we define a \emph{feature magnetization} $\mu_k=\frac{1}{N} \sum_{i=1}^N f_{ki} s_i$ and measure the feature retrieval map, plotted in panel (d) of Fig.~\ref{fig:results} for different training times.
We notice that, for the values $\alpha=0.5$ and $\alpha_D=0.1$ used, the Hebbian rule is not able to make the features locally stable ($\mu_\text{F}<1$ for $\mu_\text{I}=1$). Instead, the Daydreaming algorithm spontaneously stabilizes the features and produces large basins of attraction around the hidden features.
We do not have an explanation of why the optimal shape seems to be achieved at a finite training time and further studies are needed to answer this question.

In general, we can affirm that learning the coupling/synaptic matrix $J$ via the Daydreaming algorithm strongly enhances the stability of (and thus the possibility of retrieving) the correlated examples stored by the network, as well as the hidden features that generated those examples.

\section{Results on the MNIST dataset}
\label{sec:results_mnist}
Given the excellent performances of the Daydreaming algorithm in training the coupling matrix $J$ from correlated data, it is natural to explore its behavior when the algorithm is fed with highly structured data.
To this aim, we used the MNIST dataset \cite{lecun1998gradient}, the most widely used dataset in machine learning.
We train the model with Daydreaming on a deskewed version of the MNIST dataset: the images are deformed and rotated so that they are all centered and aligned vertically. Moreover, the images are cropped to the central area of size $14 \times 14$ pixels and made binary (with values $\pm1$) using a threshold of 86. This whole procedure was suggested in \cite{belyaev2020classification}. Finally, we split the dataset into a balanced training set (where all the ten digits have the same frequency), from which we choose examples to feed the Daydreaming algorithm, and a test set, where we keep examples that remain unseen during training.

Let us start by discussing a caveat of the Daydreaming algorithm when applied to the MNIST dataset for very high loads ($\alpha\geq100$) and a simple modification to cure it.
Consider a pair of pixels $i$ and $j$ such that $\xi_i^\mu \xi_j^\mu=1$ for all the training examples (this happens often for the pixels on the boundaries that take always the value of the background color). This very strong correlation in the data implies a very strong coupling in the trained matrix $J$. Correctly the Daydreaming algorithm tries to assign $J_{ij}=\infty$ to force pixels in $i$ and $j$ to take always the same value.
Unfortunately, if some matrix elements become extremely large, when the matrix is normalized, the remaining elements become very small and tend to vanish, thus erasing the relevant information stored in the coupling matrix.
We have solved this problem by just introducing a threshold $J_\mathrm{max}$ on the maximum absolute value that a single coupling can assume.
In the presence of a threshold value $J_\mathrm{max}$ for the couplings, it is important to properly normalize the coupling matrix in a way that the norm of $J$ does not depend on the load $\alpha$. Thus we chose to normalize the matrix obtained with the Hebb rule (i.e. the initial matrix of Daydreaming) by $P = \alpha N$ instead of $N$.
We used a threshold $J_\mathrm{max}=0.5$ for any value of $\alpha$. This proved sufficient to prevent the collapse of the matrix. We also checked that the results do not depend on the precise value of $J_\mathrm{max}$ in a range $[0.25,1.25]$. 

\begin{figure}
\centering
    \includegraphics[width=0.99\textwidth]{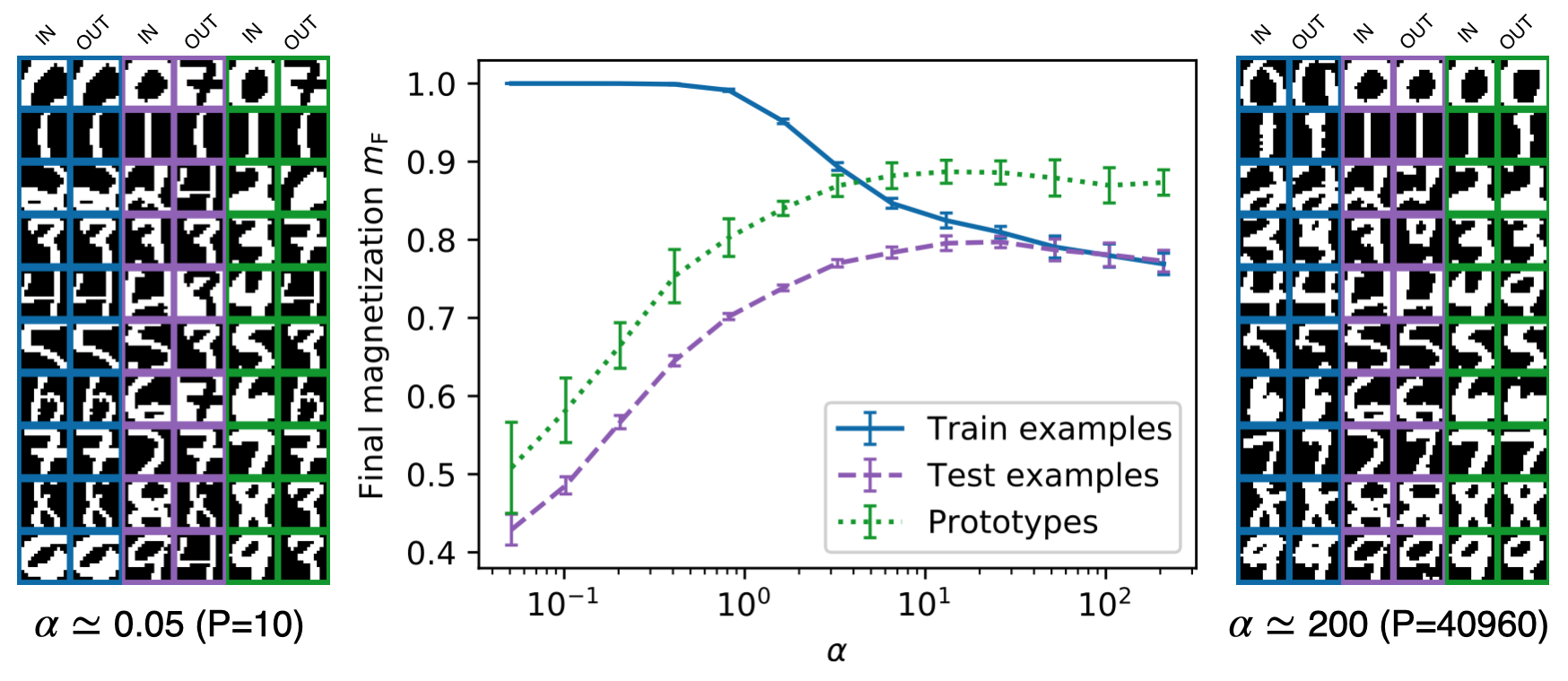}
    \caption{\textbf{Attractors close to prototypes emerge at high $\alpha$ when we train via Daydreaming on MNIST}. \emph{Central panel:} We show how the stability of examples evolves as a function of $\alpha$. Local stability is  quantified as the overlap $\mF$ between the final state of a dynamics initialized with a specific example and the example itself. When $\alpha$ is small enough, the train images are the only stable examples. When $\alpha$ is sufficiently high, the curves for train and test data coincide, indicating uniform treatment of seen and unseen images (generalization). Moreover, prototype images, defined as the averages of the training images for each digit, exhibit higher stability in this region. \emph{Lateral panels:} we show examples of the initial and final configurations of the dynamics for low and high load (left and right respectively). The colors represent training images, test images, and prototypes as in the central panel. For each couple of images, the left one is the input and the right one is the output.
    For this plot, we used $t=2^{20}$ and $\tau = 64$ for each value of $\alpha$ and a threshold $ J_\mathrm{max}= 0.5$ for the absolute value of the couplings.}
    \label{fig:results_mnist}
\end{figure}

We summarize in Fig.~\ref{fig:results_mnist} the results for the Daydreaming algorithm trained with the MNIST dataset described above.
The central panel of Fig.~\ref{fig:results_mnist} shows $m_F$ obtained starting from $m_I=1$ with examples from the training dataset (blue continuous line), examples from the testing dataset (purple dashed line), and prototypes (green dotted line). A prototype is defined as the average of all the training images corresponding to the same digit (more details are provided below).
The lateral panels provide typical examples of initial (IN) and final (OUT) configurations of the Hopfield spin update dynamics. Color codes are like in the central panel. Left and right panels are for low and high loads respectively. These results are discussed below.

\subsection{Low load}
We start commenting Fig.~\ref{fig:results_mnist} from the small $\alpha$ region.
Training examples are stable (i.e.\ $\mF\simeq1$) for $\alpha<\alpha_c\simeq 0.8$, where the critical load $\alpha_c$ is identified as the point where the magnetization of the training examples starts decreasing (solid blue curve). This result is supported by figure \ref{fig:mnist-basins} in the Appendix, where we show the retrieval maps obtained for the MNIST dataset at several values of $\alpha$: the size of the basins of attraction shrinks to zero around $\alpha\simeq 0.8$.
The model perfectly stores training examples until it reaches a threshold between $P=80$ and $160$ (corresponding to $\alpha$ between $0.4$ and $0.8$) which is comparable with recent results \cite{nouri2023eigen}, where the authors use a much more complicated attractor network to store 100 MNIST examples. Note that they use full-resolution images and therefore more parameters than our model (their $N$ is larger than ours), which means that the Daydreaming algorithm manages to store examples at a higher effective load $\alpha$.

The results achieved by the Daydreaming algorithm not only are comparable with the state-of-the-art \cite{nouri2023eigen} but also go beyond expectation. Indeed several modifications to the Hopfield model have been recently proposed based on the common idea that the standard Hopfield model is very inefficient at storing and retrieving structured data (as the MNIST dataset). Our results prove this is not the case if the coupling matrix is learned through a smart algorithm like the Daydreaming one.

In the left panel of Fig.~\ref{fig:results_mnist} we can see that the model correctly recognizes only examples from the train set (blue boxes), and it makes mistakes when presented with examples from outside the train set (purple and green boxes). This is consistent with the behavior of the Hopfield model in the storage phase.

\subsubsection{Classification}
In the regime of low load, when the model is presented with images unseen during the training, the dynamics converge to one of the stored memories (i.e. one of the training images), usually the closest one, and is not trapped in spurious minima. This can be appreciated in the left panel of Fig.~\ref{fig:results_mnist}, where all the output images correspond to one of the training images (those in the blue columns). Inspired by this behavior and by the results presented in \cite{belyaev2020classification}, where a different learning method is used, we try to use the Hopfield model equipped with the Daydreaming algorithm for digit classification. 

As a first step, we define the prototypes $\pmb{\pi}^C$ for each class $C$ (corresponding to each of the ten digits) as the binarized averages of the images in the training set within that class:
$
    \pmb{\pi}^C = \mathrm{sign} \left( \mathbb{E}_{\mu \in C} \; \pmb{\xi}^\mu \right).
$
We then proceed to train the model exclusively on these ten patterns and evaluate its performance on the test set. The label predicted by the model for a test image corresponds to the prototype that corresponds to the final state of the dynamics initialized with that test image. Instead, if the dynamics converge to a fixed point different from all prototypes (spurious pattern), then no label is assigned to that image. For the MNIST dataset, we obtain a test accuracy (i.e. a fraction of correctly assigned label) of about 67.5$\%$, which is higher than the accuracy found in \cite{belyaev2020classification} (61.5\%). More in-depth results are reported in the Appendix.
 
\subsection{High load}
If we run the Daydreaming algorithm on the MNIST database with many more training examples (large $\alpha$ regime), we observe that the training examples are no longer locally stable. Instead, the dynamics initialized on a training image end in a fixed point whose magnetization $m_F$ decreases as we increase $P$ or $\alpha$. At the same time, if we initialize the model in an example taken from the test set, we find that the final magnetization increases with $P$ or $\alpha$ (purple dashed line in Fig.~\ref{fig:results_mnist}). Interestingly enough, for $P\geq10240$ ($\alpha\geq 52.2$) the train and test magnetizations become equal, signaling that the network is acting similarly on seen and unseen examples (the generalization property).

To understand what the new attractors of the model might look like at high loads, we consider the prototypes defined above. Starting the dynamics on one of the prototypes, we observe that the curve for the final magnetization follows the same trend as the one for the test examples, increasing with $\alpha$ (green dotted line in Fig.~\ref{fig:results_mnist}). This curve is also systematically higher than that of test magnetization. This indicates that the new attractors of the model are closer to prototypes than to examples.

In the right panel of Fig.~\ref{fig:results_mnist} we see that the model does a good job at recognizing both examples from the train set (blue boxes) and examples from outside the train set (purple and green boxes). This is consistent with a generalization behavior.

\section{Conclusions, discussion, and perspectives}
\label{sec:conclusions}
To overcome a series of problems that afflicted the dreaming algorithms used to increase the storage capacity of Hopfield networks, we have designed a new learning procedure called Daydreaming.
During the training process, the Daydreaming algorithm modifies the matrix $J$ of synaptic couplings according to the rule in Eq.~(\ref{eq:learning_update}) that has the double effect of stabilizing the patterns to be stored in the network and destabilizing the spurious attractors, thus increasing the basin of attraction of the formers.  

\subsection{Effectiveness on synthetic correlated data}
Daydreaming proved to be a compact, straightforward, and streamlined algorithm, with the convergence rate notably depending only on the parameter $\tau$, which can be fixed with a large degree of freedom since the results depend on it very mildly. The Daydreaming algorithm does not suffer the problem of dreaming too much and does not require any fine-tuning, thus solving the main limitation of previously known dreaming algorithms. Moreover, it seemingly does not require any assumption on the structure of the data, as it finds large basins of attraction even for highly correlated random-features examples. 

This last point is somewhat surprising, as the classical picture in the literature is that correlation hinders retrieval \cite{amit1987information, fontanari1990storage, der1992modified, lowe1998storage, van1997hebbian}. Another surprising result is that correlated examples have larger basins of attraction than uncorrelated examples, and can be retrieved even above $\alpha=1$, which is the hard limit for uncorrelated examples \cite{gardner1988space}. 

The above observations lead us to conjecture that the Daydreaming algorithm can automatically detect and exploit the correlation in the data thus improving the storage performances.
This is supported by the fact that, in the case of synthetic data generated via the random-features model, the Daydreaming algorithm improves also the retrieval of the features.
Note that this fact is again highly non-trivial, since the update rule in Eq.~(\ref{eq:update_rule}) ignores the structure of the hidden features that generated the examples.
However, in the process of storing and reinforcing the examples the Daydreaming algorithm also learns the underlying hidden structure.

\subsection{Non-trivial attractors on MNIST}
By testing the Daydreaming algorithm on the MNIST dataset we have shown a proof of concept of how the exploitation of the hidden features might be relevant also for realistic datasets: new attractors emerge that are not exactly related to the training examples but rather to their underlying structure.

We have shown that the Daydreaming algorithm can perfectly store a large number of correlated examples (surprisingly large compared to recent results \cite{agliari2023regularization,nouri2023eigen}). This is the storage phase.

For larger loads, the storage phase finishes and one could expect a catastrophic forgetting induced by the loss of the local stability of the training examples. Indeed when the training examples are too many it is impossible for a Hopfield network to have attractors on each of them.
Surprisingly enough, in this high-load regime, a new non-trivial structure of basins of attraction emerges.

In the high-load regime, the model trained with the Daydreaming algorithm develops attractors that have a pretty large overlap ($\simeq 0.75$) with both training and testing examples (continuous blue and dashed purple curves in Fig.~\ref{fig:results_mnist}). The fact that the network responds in the same way when initialized on both training and testing examples is noteworthy, suggesting that the model approaches some sort of generalization. Moreover, the observation that class prototypes are even closer to the attractors (overlap $\simeq 0.85$, green dotted curve in Fig.~\ref{fig:results_mnist}) indicates that the emergent attractors outside the storage phase are related to the hidden structure in the data (similarly to the random-feature case).

If we interpret the training examples as noisy versions of their class prototype, the high-load energy landscape produced by the Daydreaming algorithm is reminiscent of what has been found in \cite{agliari2023regularization,alemanno2023supervised}.
A noteworthy difference between those works and the Daydreaming algorithm is that the latter is completely unsupervised, as it finds the class prototypes without using the class labels, at variance to the so-called supervised Hebb rule that instead groups data according to the class label.

Also note that, in \cite{agliari2023regularization}, the authors need to select a finite, fine-tuned value for the dreaming time. Instead, the Daydreaming procedure can be iterated indefinitely even when training on the MNIST dataset. We believe that the reason behind this difference might be related to the sampling procedures in the two algorithms: Daydreaming is essentially a non-equilibrium process. Further investigations on this point are left for the future.

To understand qualitatively how good the emergent attractors are, in the right panel of Fig.~\ref{fig:results_mnist} we have shown some examples: the output images appear as slight deformations of the inputs, but the nature of the digits appears to be preserved in most cases -- even for test examples and prototypes, which never appeared explicitly in the training phase. This is in stark contrast with the low-load regime, where the model perfectly stores train examples but fails on test examples and prototypes.
More importantly, we do not observe the dynamics converging to spurious states, since these have been mostly removed by the Daydreaming procedure. So most of the test images do converge to meaningful attractors that can be eventually used for memory retrieval and classification.
Unfortunately, we are not aware of any benchmark using the Hopfield model in this regime to compare with our results.

\subsection{Perspectives}
Given that at high loads we found indications of a rich and meaningful structure of basins of attraction, an extensive study of such structure seems promising.
In particular, it is crucial to understand the mechanism beyond the formation of this rich structure and how the Daydreaming algorithm can automatically extract hidden information from data.
Given the surprising results of Daydreaming on correlated data and its similarity to the well-known method to train Boltzmann Machines, it would be interesting to test it on more challenging datasets that would require higher-order interactions in the model.
Finally, it would be  very interesting to interpret and discuss the Daydreaming algorithm as an out-of-equilibrium process (similarly to \cite{decelle2021equilibrium,agoritsas2023explaining} in the case of Restricted Boltzmann Machines).
We leave these questions for a follow-up study.


\section{Acknowledgements}

We thank Enrico Ventura for pointing out a lot of interesting literature. CC and MN acknowledge the support of  PNRR MUR project PE0000013-FAIR. MN also acknowledges LazioInnova - Regione Lazio under the program Gruppi di ricerca 2020 - POR FESR Lazio 2014-2020, Project NanoProbe (Application code A0375-2020-36761), for its support until July 31st 2023.
RM is supported by \#NEXTGENERATIONEU (NGEU) and funded by the Ministry of University and Research (MUR), National Recovery and Resilience Plan (NRRP), project MNESYS (PE0000006) "A Multiscale integrated approach to the study of the nervous system in health and disease" (DR. 1553 11.10.2022).
FRT acknowledges financial support from ICSC – Italian Research Center on High-Performance Computing, Big Data, and Quantum Computing, funded by the European Union – NextGenerationEU.

\printcredits

\bibliographystyle{unsrt}

\bibliography{references}

\appendix
\setcounter{figure}{0}
\section{Appendix: Supplementary figures}
\counterwithin{figure}{section}

\begin{figure}[h!]    
    \centering
    \includegraphics[width=0.60\textwidth]{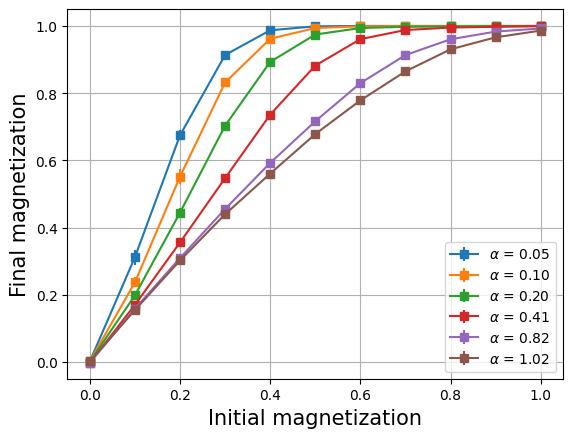}
    \caption{ 
    \textbf{The Daydreaming algorithm can perfectly store patterns from MNIST for low loads.} 
   We show retrieval maps at the end of the training procedure for various values of the load $\alpha$ for the MNIST dataset. We can observe how, up to $\alpha = 0.4$, the model develops finite attraction basins corresponding to the memories, while further increasing alpha causes the patterns to progressively lose stability until they become unstable in the high-load regime. We used $14 x 14$ images and $t=2^{20}$ for each value of $\alpha$ for this figure.}
   \label{fig:mnist-basins}
\end{figure}

\begin{table}[h!]
    \centering
    \begin{tabular}{|c|c|c|c|c|}
        \hline
        Digit & Correct class (\%)& Incorrect class (\%)& MCE & Spurious pattern (\%) \\
        \hline
            0 & 73.0 $\pm$ 3.8 & 24.8 $\pm$ 3.9 & 6 & 2.2 $\pm$ 0.8 \\
            1 & 96.1 $\pm$ 1.4 & 3.6 $\pm$ 1.1 & 0 & 0.3 $\pm$ 0.6 \\
            2 & 51.6 $\pm$ 5.2 & 45.4 $\pm$ 5.2 & 0 & 3.0 $\pm$ 1.1 \\
            3 & 66.0 $\pm$ 5.4 & 31.8 $\pm$ 5.0 & 1 & 2.3 $\pm$ 0.7 \\
            4 & 54.7 $\pm$ 8.1 & 43.8 $\pm$ 7.9 & 9 & 1.5 $\pm$ 0.8 \\
            5 & 60.2 $\pm$ 4.7 & 38.1 $\pm$ 4.5 & 3 & 1.7 $\pm$ 0.7 \\
            6 & 68.2 $\pm$ 3.6 & 29.2 $\pm$ 3.5 & 0 & 2.6 $\pm$ 0.8 \\
            7 & 79.8 $\pm$ 4.2 & 19.0 $\pm$ 4.2 & 1 & 1.2 $\pm$ 0.9 \\
            8 & 54.9 $\pm$ 5.9 & 42.9 $\pm$ 5.4 & 9 & 2.1 $\pm$ 0.9 \\
            9 & 65.7 $\pm$ 7.1 & 33.4 $\pm$ 6.9 & 7 & 0.9 $\pm$ 0.6 \\
        \hline
    \end{tabular}
    \vspace{10pt}
    \caption{Classification results for MNIST images using Daydreaming: MCE stands for Most Common Error. An image is classified as a spurious pattern if it does not coincide with any of the prototypes. Uncertainties are estimated using the empirical standard deviation between results obtained from 30 independent runs.}
    \label{tab:classification-results-Daydreaming}
\end{table}

\end{document}